\documentclass[aps,prl,twocolumn,amsmath,amssymb,floatfix,superscriptaddress]{revtex4}
\usepackage[T1]{fontenc}
\usepackage[latin1]{inputenc}
\usepackage{amsmath}
\usepackage{graphicx}
\usepackage{epsfig}
\usepackage{bm}
\def\vec{\bm}
\begin{document}

\title
{\bf A Quantum Top Inside a Bose Josephson Junction}

\author{Ingrid Bausmerth}
\affiliation{Institut f\"ur Theoretische Festk\"orperphysik,
Universit\"at Karlsruhe,
D-76128 Karlsruhe, Germany}

\author{Uwe R. Fischer$^*$}

\affiliation{Eberhard-Karls-Universit\"at 
T\"ubingen, Institut f\"ur Theoretische Physik,  
Auf der Morgenstelle 14, D-72076 T\"ubingen, Germany}

\author{Anna Posazhennikova$^{\dagger}$}

\affiliation{Institut f\"ur Theoretische Festk\"orperphysik,
Universit\"at Karlsruhe,
D-76128 Karlsruhe, Germany}


\preprint{version of \today}

\date{\today}

\begin{abstract}
We consider an atomic quantum dot confined between two
weakly-coupled Bose-Einstein condensates, where 
the dot serves as an additional tunneling channel.
It is shown that the thus-embedded atomic quantum dot is 
a pseudospin subject to an  external torque, and  
therefore equivalent to a quantum top. We demonstrate by 
numerical analysis of the time-dependent coupled evolution 
equations that this microscopic quantum top is very sensitive 
to any deviation from linear oscillatory behavior of the condensates. 
For sufficiently strong dot-condensate coupling, 
the atomic quantum dot can 
induce or modify the tunneling 
between the macroscopic condensates in the two wells.

\end{abstract}
\maketitle

In the field of ultracold atoms, whose most spectacular achievement on
relatively large scales is Bose-Einstein condensation (BEC), not only
macroscopic systems are of interest, but also to confine several or even
single atoms into optically created microtraps is becoming a potentially 
important experimental tool of what might be coined ``nanobosonics.'' 
In nanoelectronics the
control of electronic quantum dots is performed by biased conducting leads,
attached to it. In nanobosonics the role of the ``leads'' is played by
finite superfluid reservoirs of given particle number,  
which can be coupled to a particular atom by optical transitions.
Trapping and manipulating single atoms \cite{Raizen,Vuletic} opens up new
perspectives in the coherent control of quantum states, and is of relevance
for quantum computational tasks \cite{Micheli}.  

\begin{figure}[!b]\vspace*{-1.9em}
\begin{center}
\hspace*{1em}
\vspace*{-0.5em} 
\includegraphics[width=0.8\columnwidth]{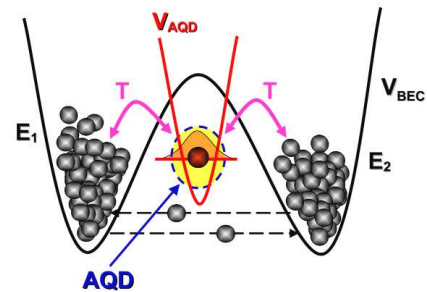}
\end{center}
\caption{\label{AQD} [Color online]  
An atomic quantum dot between two weakly-coupled
condensates, trapped in a double-well 
potential $V_{\rm \scriptscriptstyle BEC}$. 
The dot is a simple two-level system $\equiv$  
single atom present/not present and is created by the tight 
potential $V_{\rm \scriptscriptstyle AQD}$, located at the position of the 
top of the barrier.  
Atoms can be exchanged between wells
either by direct tunneling (dashed arrows) or via the dot, 
coupled to the condensates by a transfer matrix $T$.}
\end{figure}

It has recently been demonstrated by Recati 
{\it et al.} \cite{Recati} that an atomic quantum dot (AQD) 
(a single atomic two-level system), optically coupled to a 
superfluid BEC bath, can be mapped onto the spin-boson model. 
This system then exhibits a dissipative quantum phase
transition, characteristic of this model \cite{Leggett}.
Here, we study such a spin-boson model, but with a {\em time-dependent} bath: 
An AQD located inside a Bose Josephson junction (BJJ), i.e., 
a single bosonic atom coupled to two superfluid reservoirs. 
The setup under consideration is schematically 
depicted in Fig.\,\ref{AQD}. The Bose-Einstein condensate is 
trapped by the 
double-well potential $V_{\rm \scriptscriptstyle BEC}({\bm r})$.
The atom of the dot, which is in a hyperfine state different from that of the
condensate, is confined by a very tight potential $V_{\rm \scriptscriptstyle
  AQD}$, to which condensate atoms are insensitive, and which causes a large
gap for double occupation of the dot.  The coupling of the dot to the
condensates in the wells is performed in a tunable way via a Raman 
transition \cite{Recati}. 
Due to their coherent nature, the weakly-coupled condensates
exhibit quantum tunneling \cite{LeggettBEC}.  In the present paper, we
investigate the mutual influence of the induced conventional Josephson
oscillations between the two wells and the AQD, which provides an additional
tunneling channel. We demonstrate, by numerically solving the time-dependent
coupled evolution equations of AQD and condensates, that this additional
channel can in certain cases directly affect the macroscopic Josephson
tunneling.

The Hamiltonian of our system consists of three parts
\begin{equation}
H=H_{\rm cond}+H_{\rm dot}+H_{\rm coupl}. \label{H} 
\end{equation}
We will first describe these three parts in turn. 
The part $H_{\rm cond}$ characterizes the double-well trapped BEC:
\begin{eqnarray} 
H_{\rm cond} &=&\int d{\bm r}\left\{ \varPsi^{*}({\bm
  r},t)\left[-\frac{\hbar^2}{2m}\nabla^2+V_{\rm \scriptscriptstyle BEC}({\bm r})\right] \varPsi({\bm
  r},t) \right. 
\nonumber \\ 
& & \left. \qquad \qquad +\frac{1}{2} g 
|\varPsi({\bm  r},t)|^4 \right\} ,
\end{eqnarray}
where $\varPsi({\bm r},t)$ is the condensate wavefunction and 
$m$ the atomic mass.
We assume that at low energies the interparticle interaction 
is given by the usual pseudopotential $V({\bm r}-{\bm r'})=g \delta({\bm
  r}-{\bm r'})$, where $g ={4\pi\hbar^2a_s}/{m}$, and $a_s$ is the
$s$-wave scattering length. The condensate is described within  
Gross-Pitaevski\v\i\/ theory, sufficiently accurate 
at the very low temperatures we are considering \cite{Dalfovo}. 

For the present dilute bosonic gas of finite extent, the quantum tunneling
between the two wells is adequately described within a two-mode 
approximation \cite{Milburn,Smerzi};  
one expands $V_{\rm \scriptscriptstyle BEC}({\bm r})$ around each minimum, and
introduces the local mode solution, $\phi_{1,2}({\bm r})$, 
for each well separately.  
In first approximation the two modes can be
considered to be orthogonal, $\int d{\bm r}\phi_1({\bm r})\phi_2({\bm r})=0$. 
The two-mode approximation then results in the following ansatz for the total 
condensate wavefunction 
$(\varPsi_{1,2}(t)=\sqrt{N_{1,2}(t)}e^{i\theta_{1,2}(t)})$:
\begin{equation}
\varPsi({\bm r},t)=\varPsi_1(t)\phi_1({\bm r})+\varPsi_2(t)\phi_2({\bm r}). 
\label{var_wf}
\end{equation}
Since we are interested in tunneling events, i.e., in  the time 
dependence of the wavefunctions $\varPsi_{1,2}(t)$,  
it is convenient to write $H_{\rm cond}$ in the form 
\begin{eqnarray}
H_{\rm cond}&=&\sum_{i=1,2} E^0_i |\varPsi_i(t)|^2
+U_i|\varPsi_i(t)|^4
\nonumber \\ & & \qquad 
-\kappa (\varPsi_1^*(t)\varPsi_2(t)+\varPsi_2^*(t)\varPsi_1(t) ),
\end{eqnarray}
where
$E^0_{i} =\int\left[-\frac{\hbar^2}{2m}\mid\nabla\phi_{i}({\bm
    r})\mid^2+|\phi_{i}({\bm r})|^2V_{\rm \scriptscriptstyle BEC}({\bm
    r})\right]d{\bm r} $ 
are the zero-point energies in the wells $1$ and $2$, respectively, 
 the effective ``on-site'' interaction between the particles is  
given by $U_{i}=g\int|\phi_{i}({\bm r})|^4\ d{\bm r} $, and finally  
$\kappa=-\int [\frac{\hbar^2}{2m}(\nabla\phi_1({\bm r})\nabla\phi_2({\bm
    r}))+\phi_1({\bm r}) V_{\rm \scriptscriptstyle BEC}({\bm r})
\phi_2({\bm r})] d{\bm  r}$  denotes the 
coupling matrix element \cite{Smerzi}.

The Hamiltonian of the dot itself is given by
\begin{equation} 
\begin{split}
H_{\rm dot}
&=\int d{\bm r} [-\hbar \delta \hat d^\dagger({\bm r},t)\hat d({\bm r},t)  \\
&\qquad  +\frac{U_{dd}}{2}\hat d^\dagger({\bm r},t)\hat d^\dagger({\bm r},t) 
\hat d({\bm r},t)\hat d({\bm r},t)].
\label{ham_dot}
\end{split}
\end{equation}
We assume that the dot operator factorizes according to 
$\hat d({\bm r},t)= \hat d(t)\phi_d({\bm r})$, where 
$\phi_d({\bm r})$ is the spatial wave function of the 
atom on the dot normalized to unity, $\int d{\bm r} |\phi_d({\bm r})|^2=1$.  
The repulsive interaction between the dot atoms we consider to be much larger
than any other energy scales in the system, $U_{dd}\rightarrow
\infty$. The dot can then be described as a two-state system, the two states
being that an atom is or is not trapped inside the dot. 
Finally, the dot interacts with the condensate as follows 
\begin{equation} 
\begin{split}
H_{\rm coupl}&= 
g_{dc}\int d{\bm r} 
|\varPsi({\bm r},t)|^2
\hat d^\dagger ({\bm  r},t) \hat d({\bm r},t) 
\\ 
& \quad +\hbar\Omega \int \hat d{\bm r} (\varPsi^*({\bm r},t)
\hat d({\bm r},t)+{\rm h.c.}).
\end{split}
\end{equation}
Here, $g_{dc}$ is the dot-condensate 
interaction constant, and the second term describes the
coupling of the condensate atoms to the lowest vibrational state in the AQD
via a Raman transition with characteristic Rabi frequency $\Omega$. 
Spontaneous emission is suppressed by a large detuning from the excited
electronic states, which is absorbed into the effective dot energy 
$\hbar\delta$ \cite{Recati}.

To represent the evolution equations following from the Hamiltonian 
\eqref{H} in a physically transparent form, we 
introduce a new set of parameters
$ U_{id}=g_{dc}\int d{\bm r} |\phi_i({\bm r})|^2 |\phi_d({\bm r})|^2$,
$ U_{12d}=g_{dc}\int d{\bm r}
\phi_1({\bm r}) \phi_2({\bm r}) |\phi_d({\bm r})|^2$, and 
$T_{i} = \hbar \Omega\int d{\bm r} \phi_i({\bm r}) \phi_d({\bm r})$, 
with $U_{12d} = U_{12d}^*$ and $T_i = T_i^*$.    
In the single-occupation limit, 
the temporal wavefunction of the dot is just a superposition of 
singly and non-occupied states,  
$
|\varPsi_d(t)\rangle =\alpha_0(t)|0 \rangle +\alpha_1(t) |1\rangle, 
\label{wfdot}
$
where 
$|\alpha_0|^2+|\alpha_1|^2=1$. The dot
operators then 
correspond to Pauli matrices: $\hat d(t)\rightarrow \hat\sigma_-(t)$
and $\hat d^\dagger(t)\rightarrow \hat\sigma_+(t),$ introducing the 
pseudo-spin ladder operators ${\hat\sigma}_\pm = \frac12 (\hat\sigma_x \pm i\hat\sigma_y)$
in terms of the Pauli matrices 
$\hat\sigma_{x,y,z}$.
We can thus write for the coupling term 
\begin{eqnarray}
H_{\rm coupl} 
& = &  \sum_{i=1,2} \left[ U_{id}|\varPsi_i |^2
+\left( U_{12d} \varPsi_1^* \varPsi_2 + {\rm h.c.} \right) 
\right] \frac{1+\hat\sigma_z(t)}2 \nonumber \\
 &  & \qquad + \left\{ T_i \varPsi_i  \hat\sigma_+ (t) +{\rm h.c.} \right\}
\label{dot3d}
\end{eqnarray} 
One can now derive the coupled equations of motion for the condensate
\eqref{var_wf} and the spin 
$ 
\vec s(t)=\langle\varPsi_d(t)| \hat{\bm \sigma} |\varPsi_d(t)
\rangle=\langle\varPsi_d| \hat{\bm \sigma}(t) |\varPsi_d \rangle,
$ 
from the total Hamiltonian Eq.\,\eqref{H}. 
The equations for the condensate are 
\begin{eqnarray}
i\hbar\partial_t
\varPsi_1 =\left[E_1^0+U_1N_1(t)+U_{1d}n_d(t)\right]\varPsi_1  \nonumber \\
+(U_{12d}n_d(t)-\kappa)\varPsi_2 +T_{1} s_- , \nonumber \\ 
i\hbar\partial_t \varPsi_2 =
\left[E^0_2+U_2N_2(t)+U_{2d}n_d(t)\right]\varPsi_2 
\nonumber \\+(U_{12d}n_d(t)-\kappa)\varPsi_1 +T_{2} s_- ,
\label{eq_cond}
\end{eqnarray}
while the dot equations are
\begin{multline} 
i\hbar \partial_t s_- =
\left[-\hbar \delta+U_{1d}N_1(t)+U_{2d}N_2(t)+U_{12d}\varPsi_1^*\varPsi_2 \right.
\\ \left. + U_{12d}\varPsi_2^*\varPsi_1\right]s_-
-(T_1\varPsi_1+T_2\varPsi_2)s_z, 
\\
i\hbar \partial_t s_z =  2(T_1\varPsi_1^*+T_2\varPsi_2^*)s_-
-2(T_1\varPsi_1+T_2\varPsi_2)s_+.\label{eq_dot}
\end{multline} 
It is easily verified that the Eqs.\,\eqref{eq_dot} can be written 
in the vector form of a Bloch equation 
\begin{equation} 
\hbar \partial_t\vec s =\vec \omega(t) \times \vec s,
\label{svec}
\end{equation}
where the time-dependent {frequency vector} reads  
\begin{equation} 
\vec \omega (t) = 
\left(\!\begin{array}{c} 
2T_1\sqrt{N_1(t)} \cos\theta_1(t)+2T_2\sqrt{N_2(t)} \cos\theta_2 (t)\\
-2T_1\sqrt{N_1(t)} \sin\theta_1(t) -2T_2\sqrt{N_2(t)} \sin \theta_2(t) \\
-\hbar \delta+U_{1d}N_1(t)+U_{2d}N_2(t) + \omega_{12}(t) 
\cos\phi(t)
\end{array} 
\!\right)\!.
\end{equation}
where $\omega_{12}(t)=  2U_{12d} \sqrt{N_1(t)N_2(t)}$ and 
$\phi(t) = \theta_2(t) -\theta_1(t)$.  
It follows that the AQD inside the BJJ is equivalent to a quantum top. 
In the case of time-independent $\vec \omega$, Eq.\,\eqref{svec} can be solved
analytically. The presence of Josephson tunneling between the
condensates however generally results in a time-dependent 
$\vec \omega=\vec \omega (t)$, 
and the equations need to be solved numerically \cite{Ingrid}. 
\begin{figure}[!ht]
\begin{center}
\includegraphics[width=0.45\textwidth]{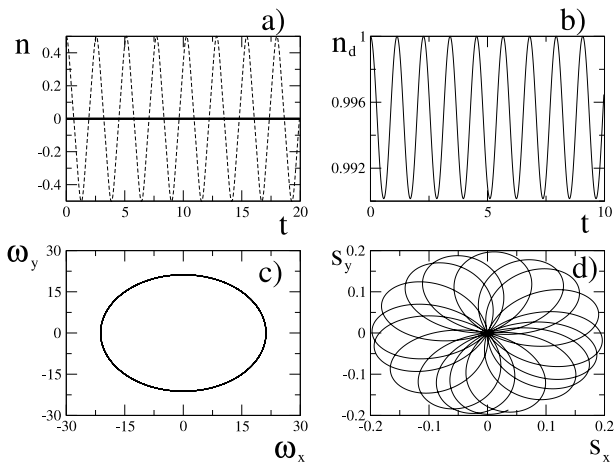}
\vspace*{-0.5em}
\end{center}
\caption{ 
  Results for weak coupling $T_{\rm rel}=0.01$ ($N_0=1500$ throughout Figs.
  2--4, as used in experiment \cite{Albiez}) and
  $\Lambda=10$.  $n(0)=0$ (Fig.2a -- solid line), $n(0)=0.5$ (Fig.\,2a --
  dashed line).  Fig.\,2b displays the dot occupation for $n(0)=0$;
  Fig.\,2c shows the precessional behaviour of $\bm \omega$ 
(in units of $2\kappa$), and  Fig.\,2d the corresponding spin nutation for $n(0)=0$.}
\label{fig2} 
\end{figure}
\begin{figure}[!ht]
\begin{center}
\includegraphics[width=0.45\textwidth]{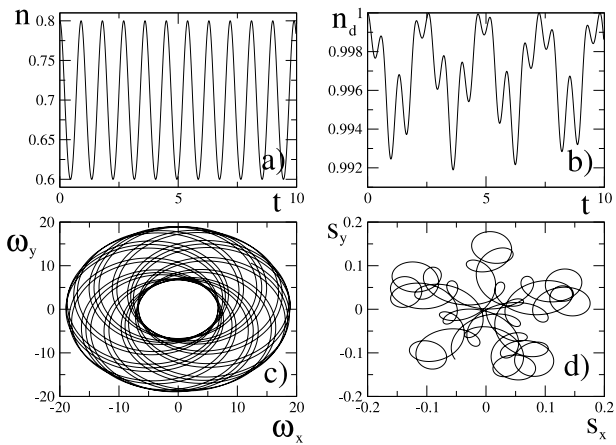}
\vspace*{-0.5em}
\end{center}
\caption{Condensate in the self-trapped MST state, $n(0)=0.8$,
  $\Lambda=10$, $T_{\rm rel}=0.01$. Relative population 
  oscillations are shown in Fig.\,3a, and the dot occupancy in Fig.\,3b. 
In Fig.\,3c, we display 
the  projection of the frequency $\vec \omega$ \eqref{svec} 
on the $x-y$ plane in units of $2\kappa$, 
 and in  Fig.\,3d the pseudospin.
}
\end{figure}

For simplicity, in what follows we consider the case
of a fully symmetric system: $E^0_1=E^0_2\equiv 0$, 
$U_1=U_2\equiv U$, $U_{1d}=U_{2d}$, $T_1=T_2\equiv T$, $U_{12d}=U_{21d}$. 
In order to compare our results with previous work on BJJ 
\cite{Smerzi}, we introduce dimensionless parameters:
$t\rightarrow 2\kappa t$, $\Lambda = U N_0 /\kappa$, 
where $N_0=N_1(0)+N_2(0)$ is the initial total number of particles in the
condensates; note that the {\em conserved} quantity is
$N_{\rm tot} = N_1(t)+N_2(t)+n_d(t)$. For numerical convenience, 
we fix the dot energy at $\hbar\delta=2\kappa$.
In addition, the interactions between AQD and condensate
$U_{1d}/(2\kappa)$ and $U_{12d}/(2\kappa)$ are assumed to be 
vanishingly small ($U_{1d}\ll U$, $U_{12d}\ll U$), and 
$\omega_z \simeq -\hbar\delta =\,$const. 
Our main parameters are then the dimensionless strength of coupling 
of dot to condensate $T_{\rm rel}=T/\kappa$, quantifying  
the relative importance of tunneling channels via dot and 
directly by conventional Josephson tunneling, 
respectively; and $\Lambda$, measuring the relative importance of 
mean-field interaction in and tunneling between the wells.

In the following, we present results for the fractional population imbalance
\begin{equation}
 n(t)=\frac{N_1(t)-N_2(t)}{N_0},
\label{nfrac}
\end{equation}
the occupation of the dot $n_d(t)$ and the trajectories of the pseudospin 
$\vec s$ on the Bloch sphere, as well as the 
projection of the frequency-vector $\vec \omega$ on the $x-y$ plane. 

We first consider the situation when the dot does not have a notable effect on
the tunneling between the wells (Figs.\,2 and 3); we fix $\Lambda=10$, and 
only change the initial condition for the particle imbalance, $n(0)$. 
The most simple situation 
is the stationary one of an initial population imbalance $n(0)=0$ and 
initial phase difference $\phi(0)=0$
(for definiteness in all figures $n_d(0)=1$, i.e., there is 
initially exactly one atom in the dot). 
These conditions result in an AQD coupled to a time-independent BEC 
\cite{Recati}, i.e., to the problem of a spin in a constant 
magnetic field, however without dissipation. 
The pseudospin generally undergoes nutation (also if we put
$n(0)=0.5$ -- Fig. \,2a -- dashed line), as shown in Fig.\,2d,
 while the vector $\bm \omega$ precesses, 
Fig.\,2c. 
However, there is an exception to this general behavior: 
For $\Lambda=1$, $n(0)=0$ 
there occurs a simple precession of the pseudospin
\cite{nutation} (not shown), 
while the occupation of the dot exhibits linear oscillations. 
For $\Lambda\neq 1$ the precession is lost, an effect 
due to the finite number of particles in the system. 

The fact that $N_{\rm tot}$ is a 
finite quantity constitutes one major difference 
to the system of a single spin coupled to superconducting leads
considered in \cite{Shnirman}. Furthermore, 
while deviation from simple precessional
behavior also occurs in that system, 
the effective fractional population imbalance 
$n(t)$ is essentially zero. 
Regimes related to large $n(t)$ 
of order unity, to be discussed below, are thus not accessible for the  
superconducting Josephson junction -- single spin system.  
In addition, the tunneling (quasi-)particles are treated as noninteracting 
in the latter case. 
Here, by contrast, including interactions between the fundamental bosons  
is crucial. In particular, as a consequence of interactions, 
and as discussed in detail in \cite{Milburn,Smerzi}, depending on
$\Lambda$ and the initial conditions, a condensate in a double-well 
potential can exhibit a novel quantum state -- macroscopic self-trapping 
(MST), successfully observed experimentally \cite{Albiez}.  MST is only 
present for the {\em self-interacting} matter waves, and is characterized 
by a nonzero time average of the population imbalance $n(t)$. The transition 
to the MST state is a gradual crossover, and we observe that our quantum top 
is very sensitive to this crossover. 
In the plots of Fig.\,2, far away from the self-trapped state, 
the coupling strength $T_{\rm rel}$ does not 
influence in a qualitative way the behavior of the quantum top. 
The pseudospin behavior 
however drastically changes as we approach MST 
It appears that the AQD is sensitive to the deviation from  
linear oscillatory behavior of the condensates occurring 
in this regime. The linear oscillation of the dot occupation
is then destroyed (Fig.\,3b), 
and the pseudospin undergoes multiple-frequency
rotations (Figs.\,3c and 3d). In the MST state, the pseudospin can thus 
behave in a rather irregular manner already for small values of the 
relative coupling $T_{\rm rel}$. 

\begin{figure}[!t]
\begin{center}
\includegraphics[width=0.45\textwidth]{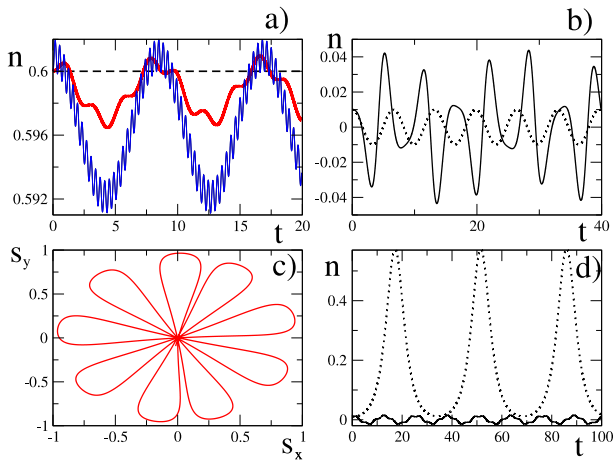}
\vspace*{-0.5em} 
\end{center}
\caption{ \label{tmatrix} [Color online] 
  Results  for a   $\pi$-junction. In Fig.4a, self-trapped MST state  
  with $n(0)=0.6$, $\Lambda=1.25$;  
  decoupled dot, $T_{\rm rel}=0$ (black dashed line); $T_{\rm rel}=0.1$ (thick
  red line), $T_{\rm rel}=1$ (thin wavy blue line); the pseudospin then
  nutates, Fig.\,4c ($T_{\rm rel} =0.1$). 
The AQD induces strong modifications both for small oscillations between 
  wells and in the MST state, Figs.\,4b and 4d; $n(0)=0.01$,  
  and $T_{\rm rel}=0$ (dotted curve), $T_{\rm rel}=1$ (solid curve). 
Weak interaction coupling, $\Lambda =0.1$ in Fig.\,4b, strongly coupled   
MST state,  $\Lambda  =1.1$, in Fig.\,4d.} 
\end{figure}

There is another potentially interesting regime, which occurs when the effect
of the dot on the tunneling between the wells becomes significant (Fig.\,4),
i.e., with increasing value of the coupling to the wells $T_{\rm rel}$.
Consider, for instance, a $\pi$-junction \cite{Smerzi}, 
$\phi(0) = \pi$, $n(0)=0.6$, $\Lambda=1.25$ 
(Fig.\,4a, dashed line).  The coupling to the dot leads to 
small oscillations between the wells (results for different $T_{\rm rel}$ are
shown in Fig.\,4a), 
and the pseudospin undergoes nutation, as apparent from Fig.\,4c.
For weak interactions (in the so-called Rabi regime \cite{LeggettBEC,Smerzi})  
and very small 
particle imbalance, the effect of the dot becomes more pronounced (Fig.\,4b). 
When the coupling to the dot is very weak,  
we observe the nutation of $\vec s$ and precession of $\vec \omega$.
Increasing $T_{\rm rel}$ leads to significant modifications of the tunneling
picture (Fig.\,4b, solid line), with strongly non-sinusoidal oscillations
of the population imbalance. Finally, we observe that,  
changing $T_{\rm rel}$ from small to large values, 
the dot can switch the BJJ from the MST state to a small
population imbalance state (Fig.\,4d).

In conclusion, we have shown that two weakly-coupled condensates, with an AQD
situated at the location of the top of the barrier between them, can exhibit
several regimes of oscillatory behavior.  The AQD behaves as a quantum top
whose behavior is very sensitive to the tunneling mode between the
condensates. Even for small couplings and stationary condensates the ``spin''
of the dot nutates, an effect due to the finite number of particles in the
system, which vanishes for an infinite system. Nutation is a characteristic
feature of the quantum top in regimes far away from the MST state. Conversely,
moving towards the self-trapped regime, we obtain strong deviations from
nutational behavior, and the quantum top motion becomes strongly irregular.
However, when the AQD itself modifies in a significant way the oscillations
between the wells, nutation can emerge also in a MST state.  Finally, the dot
can act as a switch for the BJJ from MST to small population imbalance
oscillations.

We treated the condensate on a mean-field level.
In future studies, it would be of interest to study the influence of 
condensate quantum fluctuations on the AQD \cite{Fedichev}, 
in the limiting case that the dot provides the dominant 
tunneling channel between the condensates. 

\acknowledgments
We acknowledge 
discussions with P. Coleman, M. Esch\-rig, S. Montangero, P. Pedri, A. Recati, 
G. Sch\"on, S. Shenoy, and G. Shlyapnikov.

\smallskip 
$^*$\,{\footnotesize \sf uwe.fischer@uni-tuebingen.de}     
$^\dagger$\,{\footnotesize\sf anna@tfp.physik.uni-karlsruhe.de}


\begin{thebibliography} {99}
\bibitem{nutation} The term nutation has its standard meaning of top theory 
in the present paper, i.e., nutation is generally understood to be the  
(quasi-)periodic ``wobble'' motion of a spinning top, as opposed to a simple precession.
\end{thebibliography}
\end{document}